\documentclass[aps,prl,reprint,floatfix,amsmath,amssymb,superscriptaddress,groupaddress, footnote, showkeys]{revtex4-1}

\usepackage{bbm}
\usepackage{graphicx}
\usepackage{hyperref}
\usepackage{bm}
\usepackage{printlen}
\usepackage{units}
\renewcommand{\vec}[1]{\mathbf{#1}}

\newcommand{\Trace}{\text{Tr}\,}

\renewcommand{\Re}{\text{Re}}
\renewcommand{\Im}{\text{Im}}

\hypersetup{hidelinks}

\usepackage[dvipsnames]{xcolor}

\begin{document}

\title{Entanglement entropy in lattice models with quantum metric}

\author{Alexander Kruchkov}

\affiliation{Department of Physics, Princeton University, Princeton, New Jersey 08544, USA}

\affiliation{Institute of Physics, {\'E}cole Polytechnique F{\'e}d{\'e}rale de Lausanne,  Lausanne, CH 1015, Switzerland, and Branco Weiss Society in Science, ETH Zurich, Zurich, CH 8092, Switzerland}   
 
\affiliation{Department of Physics, Harvard University, Cambridge, Massachusetts 02138, USA}

\author{Shinsei Ryu \footnotemark[1]}

\affiliation{Department of Physics, Princeton University, Princeton, New Jersey 08544, USA}

\date {\today}

\begin{abstract}
We revisit the connection between entanglement entropy and quantum metric in topological lattice systems,
and provide an  elegant and concise proof of this connection. In gapped two-dimensional lattice models with well-defined tight-binding Hamiltonians, we show that the  entanglement entropy is intimately related to the quantum metric of electronic states. 
\end{abstract}

\maketitle

\noindent
{
\textbf{{Introduction.}}}
The concept of entanglement entropy  is key in quantum information science for addressing the utility of qubit pairs, whether maximally or partially entangled \cite{Amico2008}.  Since it is feasible to convert pairs of partially entangled quantum states into fewer maximally-entangled states, the efficacy of partially entangled qubits compared to their maximally entangled counterparts is defined by their entanglement entropy \cite{Bennett1996,Calabrese2004}.  

Ground states in quantum phases of matter are defined by entanglement, setting them apart from conventional phases that are characterized by order parameters and their symmetries. These quantum states necessitate a description rooted in quantum correlations, such as topological or quantum order [4,5]. The degree of quantumness of the ground state $|\Psi \rangle$ is quantified by its entanglement entropy (von Neumann entropy),
\begin{align}
\mathcal S (\mathcal A)  = -\text{Tr}_{\mathcal A}  \, \rho_{\mathcal A} \ln \rho_{\mathcal A},
\ \ \
\rho_{\mathcal A} = \text{Tr}_{\mathcal B} | \Psi \rangle \langle \Psi | .
\end{align}
Here the system is bipartitioned into two subsystems $\mathcal A$ and $\mathcal B$, with  $ \rho_{\mathcal A} $  denoting the reduced density matrix for subsystem $\mathcal A$, derived by tracing out the subsystem $\mathcal B$ from the total density matrix $\rho = |\Psi\rangle\langle\Psi| $. The entanglement entropy $S ({\mathcal A})$ is zero for classical product states whereas it takes a non-trivial value for generic quantum-entangled systems, see e.g Refs \cite{Vidal2003, Gioev2006,Wolf2006,Klich2006,Kitaev2006,Li2008,Ryu2006adS/CFT}.

Previous work 
%by Ryu and Hatsugai 
\cite{Ryu2006} has established  hidden relations between topological properties of a system, quantum metric in 1D, and bounds on entanglement entropy. In particular, for one-dimensional systems  
Ref.\ \cite{Ryu2006}
%\textit{Ryu-Hatsugai formula} 
provides the exact relationship between the  entanglement entropy $\mathcal S^{\rm (1 D)}$ the Berry phase $\gamma$ in the case of the flat-band limit,
\begin{align}
\mathcal S^{\rm (1 D)} = - \frac{\gamma}{\pi} \ln \frac{\gamma}{2 \pi} - \frac{2 \pi - \gamma}{\pi} \ln \frac{2 \pi - \gamma}{2 \pi}	.
\end{align}
Moreover, for the ground states with chiral symmetry, entanglement entropy has lower bound $S^{\rm (1 D)} \geq \ln 2$ (per boundary), i.e., the value of entanglement entropy for a maximally-entangled qubit pair \cite{Ryu2006}. Such relationships in one-dimensional systems substantiate the existence of definitive bounds on entanglement entropy. However,  extending this reasoning to higher dimensions, including recently discovered two-dimensional topological phases, remains an open research question.

On the other hand, quantum metric is a fundamental concept that governs distance  between quantum states in the projective Hilbert space \cite{Provost1980}. If Bloch basis $| u_{n \vec k} \rangle $ is well defined, the solid-state quantum-geometric tensor is defined  as \cite{Matsuura2010}
\begin{align}
\mathfrak G^{}_{ij} (\vec k)	 = \langle \partial_{i} u_{n \vec k}| ( 1- \mathcal P_{n \vec k } )  |  \partial_{j} u_{n \vec k}\rangle. 
\end{align}
 The projector $ \mathcal P_{n \vec k }  \equiv   | u_{n \vec k}\rangle    \langle u_{n \vec k}| $ ensures  gauge-invariance of observables \cite{Matsuura2010}. 
The real part of $\mathcal G_{ij} = \Re \mathfrak G_{ij}$ is the Fubini-Study metric describing the geometry of the electronic bands, while the imaginary part $\mathcal F_{ij} = -2 \Im \mathfrak G_{ij}$ is the Berry curvature reflecting the topology of quantum states. 
Although used in earlier applications of  quantum information, particularly through the  Fischer information metric  related to Fubini-Study metric \cite{Anandan1990, Braunstein1994, Facchi2010}, the connection of the quantum metric to the standing question of entanglement entropy bounds remains obscure \cite{Paul2024}.

Hence it is intriguing to seek for \textit{connection between two a priori unrelated constructions}---entanglement and quantum metric---which is the main result of this Letter.  We find that for the \textit{lattice  models with well-defined tight-binding} such connection does indeed exist, and, depending on partition, can be expressed through quantum metric invariants.

\

{
\textbf{{Methodology.}}} We here make use of the formalism introduced by Jin and Korepin \cite{Jin2004}, and further developped in the works of Calabrese, Mintchev, and Vicari \cite{Calabrese2011,Calabrese2011b, Calabrese2012}; refer also to Paul \cite{Paul2024}. Within this formalism, the system is bipartioned into  entanglement area ${\mathcal A}  $  and the rest.  The total entanglement entropy can be calculated as
\begin{align}
\mathcal S_{\alpha} ( {\mathcal A} ) =  \sum_{k} s_{\alpha} (\lambda_{ k})
\label{entan-entropy}
\end{align}
where Renyi entropies are given by
\begin{align}
s_{\alpha} ( \lambda ) = \frac{1 }{1- \alpha} \ln [ \lambda^{\alpha} +  (1-\lambda)^{\alpha}], 
\label{Renyi-entropy}
\end{align}
following limit $\alpha \to 1$ to obtain von Neumann entropy.  Here $\lambda_{k}$ are eigenvalues of the \textit{overlap matrix}  \cite{Klich2006, Calabrese2011}. We here consider a  two-orbital band  insulator at half-filling. 
For clarity, we consider a 2D material stripe, partitioned by a straight cut at $L_{\mathcal{A}}$, with momentum along the partition boundary denoted by $k$. The overlap matrix is given by 
\begin{align}
\mathcal W_{ k_1  k_2}	 =   \int  \limits_{\vec x \in \mathcal A} d \vec x  \,  \psi^{*}_{ k_1} (\vec x) \psi^{}_{ k_2 } (\vec x). 
\end{align}
The overlap matrix $\mathcal W_{k_1 k_2}	$ encodes information on the overlap of Wannier orbitals, i.e. entanglement in the system. On the other hand, overlap between electronic orbitals can be characterized by the quantum metric \cite{Marzari1997}. 
Therefore, we need to investigate the spectrum of the overlap matrix and compare it with Marzari-Vanderbilt functional. We show below that these two quantities are related.

	\begin{figure}[b]
\includegraphics[width=1.0 \columnwidth]{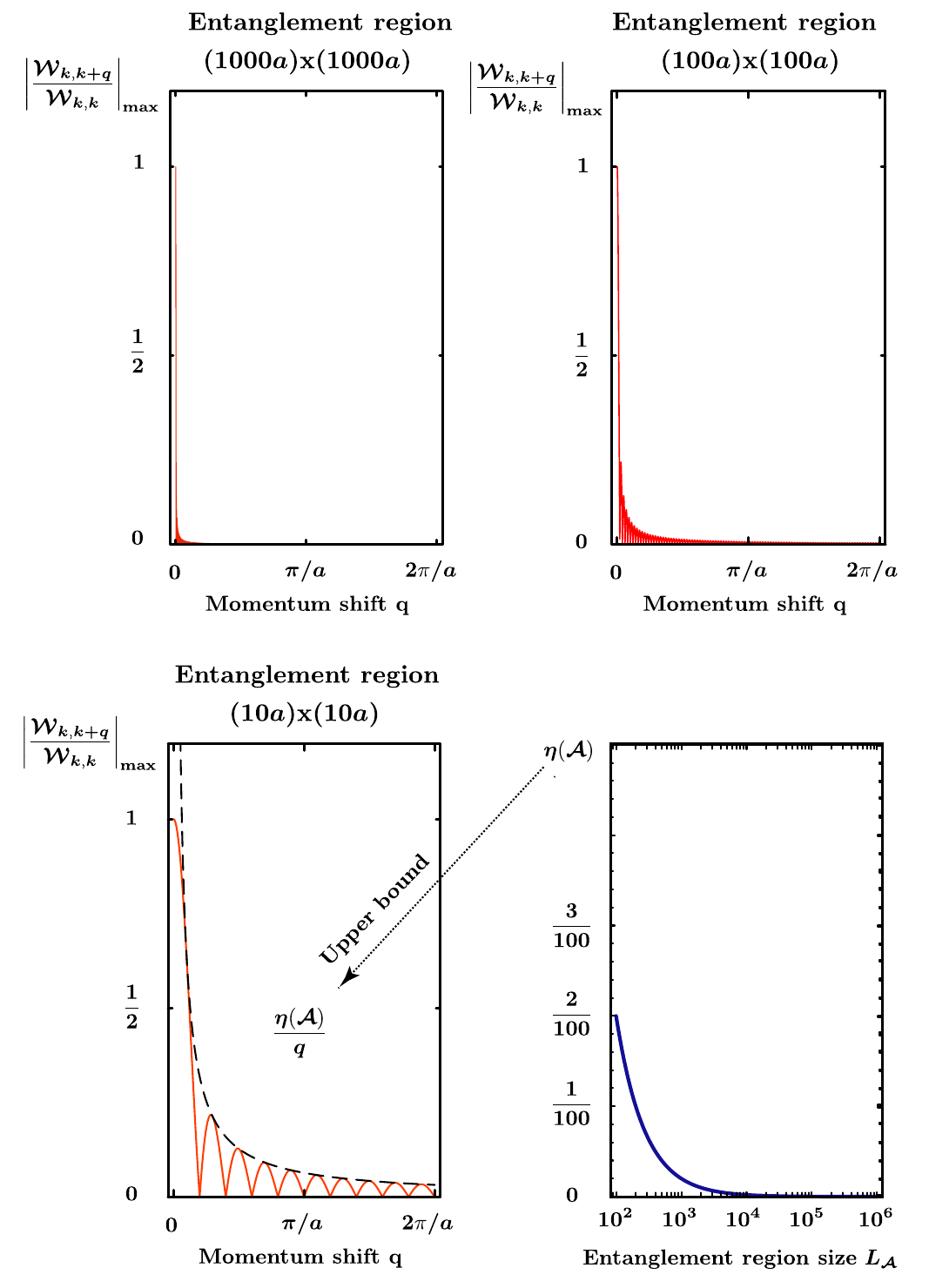}
	\caption{Upper bounds on overlap matrix elements show asymptotical decay of $1/q$, with a prefactor which is significantly suprressed for the mesoscopic samples. For sample $1000 a$x$1000 a$   (here $a$ is lattice constant), the decay in matrix elements is abrupt and focused around $q \approx 0$. Larger entanglement regions provide significantly supressed values for $q \ne 0$ terms.  }
	\end{figure}

We examine two-dimensional topological lattice models with well-defined tight-binding  description. The tight-binding description and the underlying lattice are conceptually important for our derivation. 
In what follows below, we consider solid-state materials with Bloch basis well-defined, 
$
\psi_{\vec k} (\vec x) =  e^{i \vec k \vec x} u_{ \vec k} (\vec x) . 
$
In tight-binding formalism, the overlap matrix is 
\begin{align}
\mathcal W^{}_{ k_1  k_2}	 &=  \sum_{x \in A} \sum_{j} e^{i (k_2 - k_1) x} u^*_{k_1} (j) u_{k_2} (j) 
\\
&=  \sum_{x \in A} e^{i (k_2 - k_1) x}  \langle u_{k_1}  | u^{}_{k_2}  \rangle .
\label{overlapmatrix2}
\end{align} 
Introducing now $\chi_q = \sum_{x \in A} e^{i q x}$, the overlap matrix reads 
 \begin{align}
\mathcal W^{}_{ k_1, k_2}	=  \chi_{k_2-k_1}     \langle u_{k_1}  | u^{}_{k_2}  \rangle .
\label{overlap3}
\end{align}
For notational simplicity, we discretize momenta in the square-lattice Brillouin zone. Extending this approach to other lattice symmetries is straightforward.

{\textbf{Derivations for entanglement entropy}}.  We start with investigating asymptotical properties of off-diagonal elements of the overlap matrix \eqref{overlap3}. 
The structure of matrix elements \eqref{overlap3} is such that it rapidly decays in mesoscopic samples, oscillating between two strict bounds, (Fig. 1), 
\begin{align}
0 \leq  \left | \frac{ \mathcal W_{ k,  k +  q}  }{ W_{ k,  k } }  \right|  \leq \frac{\eta (\mathcal A) }{|q|} ,	  \ \ \ \eta (\mathcal A) \ll 1.  
\end{align}
hence the main contribution comes from small $ q \approx 0$. 
Note that the upper bound asymptote $a (\mathcal A)/q$ depends on the entanglement area $\mathcal A$, and is small for physically relevant entanglement regions of mesoscopic sizes (see Fig. 1), In fact,  $\eta (\mathcal A)$ decays as approximately $1/ L_{\mathcal A}$, inversely to the size of entanglement region, reaching values for around $\eta \sim 10^{-3}$ for physically-relevant sizes $L_{\mathcal A} \sim 10^3$ (lattice size $a=1$) and $\eta \sim 10^{-6}$ for $L_{\mathcal A} \sim 10^{6}$ sampling. Hence, the off-diagonal matrix elements decay sufficiently fast in mesoscopic and macroscopic samples, and the overlap matrix can be effectively truncated \cite{Mirlin1996}.  Therefore, this problem can be addressed from the perspective of  "tight-binding in momentum space" \cite{Matsuura2010}, truncating the "momentum hopping" to the first nearest neighbors.  To treat problem analytically, we first truncate the overlap matrix to tridiagonal form,
\[
\begin{footnotesize}
\setlength{\arraycolsep}{1.5 em} 
\renewcommand{\arraystretch}{4} 
\begin{pmatrix}
\ddots & \ddots & 0 & 0 & 0  \\
\ddots & \mathcal W_{k-q, k-q} & \mathcal W_{k-q, k} & 0 & 0  \\
0 & \mathcal W_{k, k-q} & \mathcal W_{k, k} & \mathcal W_{k, k+q} & 0  \\
0 & 0 & \mathcal W_{k+q, k} & \mathcal W_{k+q, k+q} & \ddots \\
0 & 0 & 0 & \ddots & \ddots \\
\end{pmatrix}
\end{footnotesize}
\]
On the main diagonal we have $\chi_0$, the secondary diagonals have properties $\mathcal W^{\dag}_{i, i+1} = \mathcal W_{i+1, i} $, hence it is a Hermitian matrix with real eigenvalues.

	For the reasons which will become clear shortly,  we further consider  the \textit{squared} overlap matrix $\mathcal W^2$. While it does not seem to be possible to find its eigenvalues in the closed form,  we can write down its diagonal elements. Its \textit{diagonal elements} are given by relatively simple expression
\begin{align}
\left [ \mathcal W^2_{\text{tri}} \right]_{kk} = 	\chi_0^2  & + | \chi_q |^2  \,  |\langle u_k | u_{k+q} \rangle |^2   \nonumber 
\\ 
& + | \chi_q |^2 \, |\langle u_k | u_{k-q} \rangle |^2  ,
\label{diag-squared}
\end{align}
which will become handy for evaluating entanglement entropy.

We  further consider von Neuman entropy corresponding to the eignevalue of $\mathcal W$, 
\begin{align}
s_{\alpha \to 1}  (\lambda)  \approx - \lambda \ln \lambda - (1- \lambda) \ln (1- \lambda )   	.
\end{align}
	  With a good accuracy, this  formula can be approximated as (see Fig. 2 ) 
\begin{align}
s  (\lambda) \approx 4 \ln (2) \ \lambda (1 - \lambda )   . 
\end{align}
Thereafter, the trace can be approximated as 
\begin{align}
S \simeq 4 \ln 2 \left[  \Trace {(\mathcal W)} - \Trace (\mathcal W^2)  \right] .
\label{charge fluct}
\end{align}
Substituting now formula \eqref{diag-squared} in \eqref{charge fluct}, we obtain
\begin{align}
S_{\text{tri}}\simeq  
\frac{4 \ln 2}{N} \sum_{q \ne 0} \chi_0 -  	\chi_0^2  & - | \chi_q |^2   |\langle u_k | u_{k+q} \rangle |^2  
\nonumber 
\\ &  - | \chi_q |^2   |\langle u_k | u_{k-q} \rangle |^2 
\end{align}
here $N$ is the dimension of the matrix, given by discretization scheme $k_j = \frac{2 \pi}{a}\frac{j}{N}$, $q = \Delta k = \frac{2 \pi}{a N}$. 
We now use the properties of quantum metric \cite{Matsuura2010}
\begin{align}
	 |\langle u_k | u_{k+q} \rangle |^2    = 1 - q^2  \mathcal G_k  + \mathcal O (q^4)
\end{align}
And thus we obtain entropy as a function of quantum metric, 
\begin{align}
S_{\text{tri}} \simeq \frac{ 4 \ln 2}{N} \sum_{  k \in \text{BZ} } \chi_0   - \frac{ 4 \ln 2}{N} \sum_{k \in \text{BZ} } ( \chi_0^2 +| \chi_q |^2) 
\nonumber
\\ + \frac{ 4 \ln 2}{N} \sum_{k \in \text{BZ}} q^2 |\chi_q|^2 \left[ \mathcal G_k + \mathcal G_{k-q}  \right]
\end{align}
Therefore, even in tridiagonal truncation, we observe that the entanglement entropy contains information about the quantum metric,
\begin{align}
\mathcal S_{\text{tri}} \simeq \text{Const} 
+ \frac{8 \ln 2}{N} \sum_{k \in \text{BZ} } q^2 |\chi_q|^2  \mathcal G_k . 
\end{align}
{We will show below that this trend preserves for "longer-range" momentum-space hopping.}

	\begin{figure}[t]
	\includegraphics[width=0.85 \columnwidth]{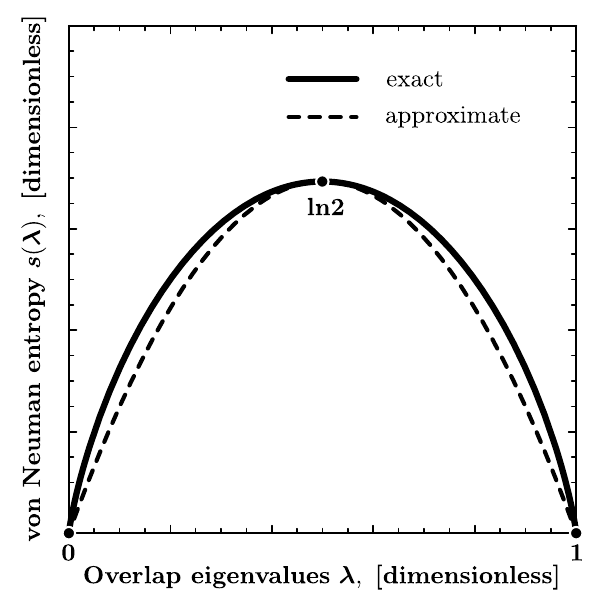}
	\caption{Approximations used to calculate the entanglement entropy.  }
\end{figure}

\textit{Pentadiagonal matrix.} To enhance our understanding, we consider next-nearest-neighbor (NNN) tight-binding hopping in the overlap matrix, resulting in a pentadiagonal matrix,---a matrix with five nonzero diagonals. Similar arguments apply, beginning with the matrix elements:
\begin{align}
\left[ \mathcal W^2_{\text{penta}}  \right ]_{kk}  = \chi_0^2 + |\chi_q |^2  |  \langle u_k | u_{k+q} \rangle |^2+  |\chi_q |^2  |  \langle u_k | u_{k-q} \rangle |^2
\nonumber
\\
+ |\chi_q |^2  |  \langle u_{k+q} | u_{k+2q} \rangle |^2 + |\chi_q |^2  |  \langle u_{k-q} | u_{k-2 q} \rangle |^2
\end{align}
Using Eqs. (14) and (16), the entanglement entropy is then given by 
\begin{align}
\mathcal S_{\text{penta}} \simeq \text{Const} 
+ \frac{4 \ln 2}{N} \sum_{k \in \text{BZ} } & q^2 |\chi_q|^2 
\nonumber
\\
\times & \left[ \mathcal G_{k+q} + \mathcal G_k +  \mathcal G_{k-q}  + \mathcal G_{k-2q} \right] . 
\end{align}
{We see that in this case as well, the entanglement entropy depends on quantum metric.}

\textit{Septadiagonal matrix.} ---{Hopping further}, we consider next-next-nearest-neighbor (NNNN) tight-binding hopping in the overlap matrix, resulting in a \textit{septadiagonal matrix},---a matrix with seven nonzero diagonals. The same principles apply, starting with the matrix elements:
\begin{align}
\left[ \mathcal W^2_{\text{septa}}  \right ]_{kk}  = & \chi_0^2 + |\chi_q |^2  |  \langle u_k | u_{k+q} \rangle |^2+  |\chi_q |^2  |  \langle u_k | u_{k-q} \rangle |^2
\nonumber
\\
+ & |\chi_q |^2  |  \langle u_{k+q} | u_{k+2q} \rangle |^2 + |\chi_q |^2  |  \langle u_{k-q} | u_{k-2 q} \rangle |^2
\nonumber 
\\
+ & |\chi_q |^2  |  \langle u_{k+2 q} | u_{k+3q} \rangle |^2 + |\chi_q |^2  |  \langle u_{k-2q} | u_{k-3 q} \rangle |^2. 
\end{align}
Using Eqs. (14) and (16), the entanglement entropy is then given by 
\begin{align}
\mathcal S_{\text{penta}} & \simeq \text{Const} 
+ \frac{4 \ln 2}{N} \sum_{k \in \text{BZ} }  q^2 |\chi_q|^2 
\nonumber
\\
& \times  \left[ \mathcal G_{k+2 q}+ \mathcal G_{k+q} + \mathcal G_k +  \mathcal G_{k-q}  + \mathcal G_{k-2q} + \mathcal G_{k-3q} \right] . 
\end{align}
{Below we generalize the quantum metric contribution to arbitrary hopping range.}

\textit{General case.}---Consider now a general case of tight-binding range $\Lambda$ which leads to $(2 \Lambda+1)$-diagonal matrix.  The diagonal elements of the squared matrix are given by
\begin{align}
	\left[ \mathcal W^2_{{\Lambda}}  \right ]_{kk}    = & \chi_0^2 +  \sum_{Q = n q}^{|n|<\Lambda/2} |\chi_q |^2  |  \langle u_k | u_{k+Q} \rangle |^2 
	\end{align}
The entanglement entropy is then given by
 \begin{align}
\mathcal S & \simeq \text{Const} 
+ \frac{4 \ln 2}{N} \sum_{k \in \text{BZ} }  q^2 |\chi_q|^2 
 \sum_{Q = nq }^{|n<|\Lambda/2|} \mathcal G_{k+Q} 
 \\
& \simeq \text{Const} 
+ [2 \ln (2) \eta^2 (\mathcal A) ]\frac{1}{N} \sum_{k \in \text{BZ} } 
 \sum_{Q = nq }^{|n<|\Lambda/2|} \mathcal G_{k+Q} 
 \end{align}
where we have used that $|\chi_q|^2 |q|^2 \approx \frac{1}{2} \eta^2 $ upon averaging over fast fluctuations (see Eq. 10 and Fig. 1) .

In practice, the quantum metric is highly nonhomogeneous, typically peaking at high symmetry points (see, e.g., Fig. 5 in Ref. \cite{Guan2022}). Therefore, it is sufficient to consider only a finite number of points around these peaks to evaluate Eq. (25).   One may, but not obliged to set $\Lambda$ to large values, so that the integration covers the entire BZ. In this case one obtains
 \begin{align}
\mathcal S \simeq \text{Const} 
+ 2 \ln (2) \ \eta^2 (\mathcal A) \sum_{k  } 
 \mathcal G_{k}. 
\end{align}
In homogeneous system, $\mathcal G_{xx} = \mathcal G_{yy} $, so one can write 
 \begin{align}
\mathcal S \simeq \text{Const} 
+ 2 \ln (2) \ \eta^2 (\mathcal A) \sum_{k  } 
\Trace  \mathcal G_{ij} (k).  
\end{align}
This is the direct relation with Marzari-Vanderbilt invariant.

\textit{Arbitrary partition region}. The results described above can be in large extended to codimension-one partitioning of arbitrary shape.  Indeed, in this case we have a band matrix with finite number of nonzero diagonals $\Lambda$. We use 
\begin{align}
\left[\mathcal W^2 \right]_{ii}=\sum_{j} \mathcal W_{ij} \mathcal W_{ji} = \sum_{j} |\mathcal W_{ij} |^2, 
\end{align}
where $i,j$ label the 2D momentum discretisation, and we have used Hermitian symmetry of $\mathcal W$. 
Using now $\mathcal W_{\vec k, \vec k'} = \chi_{\vec k'-\vec k} \langle u_{ \vec k} | u_{\vec k'} \rangle $, we hence obtain 
\begin{align}
\left[\mathcal W^2 \right]_{\vec k \vec k} = \sum_{\vec k'} | \chi_{\vec k'-\vec k}  |^2 |\langle u_{ \vec k} | u_{\vec k'} \rangle| ^2. 
\label{cod-zero}
\end{align}
Further we take into account that the quantum metric in real materials is peaked around high-symmetry points of the BZ (for example, in twisted bilayer graphene it is the $\Gamma$ point \cite{Guan2022}). In this case, it is sufficient to expand \eqref{cod-zero} in the vicinity of the high-symmetry point(s), 
\begin{align}
\left[\mathcal W^2 \right]_{\vec k \vec k}  
& = \sum_{\vec Q} | \chi_{\vec Q}  |^2  |\langle u_{ \vec k} | u_{\vec k + \vec Q} \rangle| ^2 
\\
& \simeq  \sum_{\vec Q} | \chi_{\vec Q}  |^2 (1 - \mathcal G_{ij} (\vec k) Q_i Q_j) 
\label{cod-zero}
\end{align}
Thus, the entanglement entropy has quantum metric contribution, 
 \begin{align}
\mathcal S_{\text{geo}} \simeq 2 \ln (2) \ \eta^2 (\mathcal A) \sum_{\vec k  } 
\Trace  \mathcal G_{ij} (\vec k)  .
\end{align}
{This formula represents quantum-geometric contribution to entanglement entropy. The trace of $\mathcal G_{ij} (\vec k)$, known as the Marzari-Vanderbilt invariant \cite{Marzari1997} of the quantum metric, is bounded below by the Chern number in the case of a Chern insulator. }

\

{\textbf{Relation to particle number fluctuations}}.
Klich pointed out that entanglement entropy must have lower bounds related to particle number fluctuations \cite{Klich2006}. Below, we consider a generalized fluctuation operator $\mathcal F_ {\alpha }$ (here $\alpha \in \mathbb Z$)  of the form 
\begin{align}
\mathcal F_ {\alpha } =  \left	\langle 
 \frac{\partial^{\alpha }  n (\vec x ) }{\partial \vec x^{\alpha}}  
 \frac{\partial^{\alpha }  n (\vec x ) }{\partial \vec x^{\alpha}}
  \right \rangle 
  -  \left|  \left	\langle 
 \frac{\partial^{\alpha }  n (\vec x) }{\partial \vec x^{\alpha}}  
  \right \rangle  \right|^2 .
  \label{fluct}
\end{align}
After applying  Fourier transform, we get
\begin{align}
\mathcal F_ {\alpha } =  \sum_{\vec q \ne 0}  |\vec q|^{2 \alpha}  \langle 
 n_{\vec q }  n_{-\vec q } \rangle ,
\end{align}
where $n_{\vec q}$ is Fourier-transform of the particle density operator, which in the Bloch basis reads (see, e.g., \cite{Parameswaran2013}),
\begin{align}
n _{\vec q} = \sum_{n,m}  \sum_{\vec k} \langle u_{n,  \vec k + \vec q} | u_{m ,\vec k } \rangle   c^{\dag}_{n, \vec k + \vec q} 	c^{}_{m,\vec k } .
\end{align}
Substituting this into Eq. \eqref{fluct}, and using Wick's theorem, we obtain
\begin{align}
\mathcal F_ {\alpha }  =  \sum_{\vec q} |\vec q|^{2 \alpha}   
\langle  u_{n, \vec k + \vec q} | u_{m, \vec k } \rangle 
\langle  u_{m, \vec k } |  u_{n, \vec k + \vec q} \rangle   \nonumber
\\
\times  \langle c^{\dag}_{n, \vec k + \vec q}   c^{}_{n, \vec k + \vec q} \rangle 
 \langle c^{}_{m, \vec k }   c^{\dag}_{m, \vec k } \rangle ,
 \label{fluct2}
\end{align}
We can further simplify this expression by noting that, for noninteracting fermions at equilibrium, $ \langle c^{\dag}_{n\vec k }   c^{}_{n\vec k } \rangle 
 = f_{n \vec k} $, and $  \langle c^{}_{n\vec k }   c^{\dag}_{n\vec k } \rangle 
 = 1- f_{n \vec k} $. Introducing  the projector on the occupied states 
\begin{align}
\tilde {\mathcal P}_{\vec k} =  \sum_{n } f_{n \vec k }    \ 
  |  u_{n, \vec k } \rangle  \langle u_{n, \vec k }  |  ,
\end{align}
and the antiprojector, 
\begin{align}
\tilde { \mathcal Q}_{\vec k} =  1-  \sum_{n } f_{n \vec k }    \ 
  |  u_{n, \vec k } \rangle  \langle u_{n, \vec k }  | ,
\end{align}
Eq. \eqref{fluct2} now reads as 
\begin{align}
\mathcal F_ {\alpha } =  \sum_{\vec k, \vec q} |\vec q|^{2 \alpha}   \ 
\Trace  \left[ \tilde {\mathcal P}_{\vec k}  \tilde {\mathcal Q}_{\vec k+ \vec q} \right]. 
 \label{fluct3}
\end{align}
In this form, the expression \eqref{fluct3} resembles the \textit{Marzari-Vanderbilt invariant} $\Omega_I$,  which is the integrated trace of the quantum metric in its discretized form \cite{Marzari1997}  
\begin{align}
\Omega_I =  \sum_{\vec k, \vec q} w_{\vec q}   \ 
\Trace  \left[ \tilde {\mathcal P}_{\vec k}  \tilde {\mathcal Q}_{\vec k+ \vec q} \right]
 \label{MVI}
\end{align}
where $w_{\vec q}  \propto /q^2$ and its prefactor depends on lattice symmetry \cite{Marzari1997}.  Comparing now Eq. \eqref{fluct3} with \eqref{MVI}, we conclude that the fluctuations of the excess particle number in the entanglement area $N_{\mathcal A} (t) = \int_{\mathcal A} d \vec x \, n(\vec x, t)$ are related to quantum metric, 
\begin{align}
\mathcal F_ {\text{-}1 } \propto \sum_{\vec k } \Trace \mathcal G_{ij } (\vec k) .
 \end{align}
Following Klich's argument \cite{Klich2006}, this establishes a  bound on entanglement entropy.

\

\noindent{
\textbf{{Discussion.}}}   We have derived a concise formula relating quantum-geometric contributions to entanglement entropy, 
 \begin{align}
\mathcal S_{\text{geo}} \simeq 2 \ln (2) \ \eta^2 (\mathcal A) \sum_{ \vec k } 
\Trace  \mathcal G_{ij} (\vec k).  
\label{result}
\end{align}
The connection between entanglement entropy and the trace of the quantum metric can be intuitively understood through the overlap of electronic orbitals, expressed through invariants of quantum metric \cite{Marzari1997}. Greater orbital overlap leads to higher entanglement in Eq.  \eqref{result}.
 It is also known that entanglement entropy is related to quantum noise in the system \cite{Klich2009}. Recent studies have shown that quantum noise in topological systems is directly linked to quantum metric invariants \cite{KRYU2023, Neupert2013}. Thus, the connection between entanglement entropy and quantum metric invariants is well-founded.

In conclusion, our analytical results demonstrate that entanglement entropy in topological lattice models is intrinsically linked to the underlying quantum metric.
Extending this framework to Fractional Chern Insulators could provide valuable insights.

\

\textit{Acknowledgements.} {This project is supported by the Swiss National Science Foundation, grant CRSK-2\_221180 and program IZSEZ0\_223932 "Quantum entanglement, quantum metric, and correlated electrons"}. A.K. is supported by the Branco Weiss Society in Science, ETH Zurich. S.R. is supported by 
a Simons Investigator Grant from the Simons Foundation (Award No. 566116). This work is supported by the Gordon and Betty Moore Foundation through Grant GBMF8685 toward the Princeton theory program.

 \

 \bibliography{Refs}

\end{document}